# Three-dimensional nanoimprint lithography using two-photon lithography master samples

Rebecca Saive[1*], Colton R. Bukowsky[1], and Harry A. Atwater[1]

[1] Department of Applied Physics and Material Science, California Institute of Technology, Pasadena, CA 91125, USA

## Abstract

We demonstrate three-dimensional (3-D) nanoimprint lithography using master samples initially structured by two-photon lithography. Complex geometries like micro prisms, micro parabolic concentrators, micro lenses and other micrometer sized objects with nanoscale features are three-dimensionally fabricated using two-photon lithography. Stamps made out of polydimethylsiloxane are then cast using the two-photon lithographically structured samples as master samples. Hereby, expensive serial nano 3-D printing is transformed into scalable parallel 3-D nanoimprint lithography. Furthermore, the transition from two-photon lithography to imprint lithography increases the freedom in substrate and ink choice significantly. We demonstrate printing on textured surfaces as well as residue-free printing with silver ink using capillary action.

## Introduction

Three-dimensional (3-D) micro- and submicrometer structures play an increasingly important role in many modern fields of technology such as (opto)electronics[1, 2] and optics[3, 4]. Conventional structuring methods such as deposition and etching are constrained to a limited number of geometries. Furthermore, extensive process development is needed in order to fabricate objects with the desired shape and to modify existing geometries. Until recently, there have been few options for directly creating 3-D structures, as opposed to tweaking process parameters and using multiple processing steps to achieve a desired geometry.

More sophisticated 3-D micro structures can be prepared using focused ion beam (FIB) or two-photon lithography processing. In two-photon lithography the photoresist is polymerized at the focal point of a laser leading to the ability to write 3-D structures with feature sizes down to 150 nm[5]. Therefore, two-photon lithography is also referred to as nano 3-D printing. Similar to other 3-D printing techniques the desired shape can be drafted with a computer assisted design (CAD) program and the obtained file can be converted into a print file. Shape modifications can be easily performed with few changes in the process parameters. Two-photon lithography has led to numerous scientific achievements[6] ranging from nano and micro mechanics[7-10] to nano and micro optics[3, 4, 11, 12].

However, widespread application will be limited by insufficient scalability and constraints in substrate and resist choice. Scalability is limited by the writing time (serial print method) while the substrate choice is limited to smooth, non-reflective surfaces with a refractive index different from the photo resist material. Furthermore, the development of photoresist is sophisticated and adding functionalities such as specific optical or electrical properties is non-trivial. A more scalable, flexible and robust printing method is represented by nanoimprint lithography[13]. In nanoimprint lithography a master structure is reproduced through preparation of a polydimethylsiloxane (PDMS) stamp and subsequent printing using a partially reacted sol or an ink. Different inks and sols can provide a wide variety of mechanical, thermal, optical and electrical properties[14]. Furthermore, the limitations of the substrate are low and even printing on textured surfaces can be performed.

We show a printing process that combines the design flexibility of two-photon lithography with the scalability of imprint lithography by using two-photon lithographically prepared geometries as master structures for stamp fabrication[15]. Replicas are a virtually perfect match of the originals and the shape quality is solely determined by the master structure.

Furthermore, we demonstrate a high fidelity ink process for residue free printing of silver ink lines using capillary action[16, 17].

## Methods

First 3-D structures that later will be used as master samples are prepared by two-photon lithography. In two-photon lithography polymerization of the photoresist occurs in the focus point of a laser by a two-photon absorption process[18]. The lateral laser position as well as the focus height of the laser can be varied such that 3-D structuring is achieved. We used a Nanoscribe Photonic Professional GT[19] and the photoresist IP-Dip (by

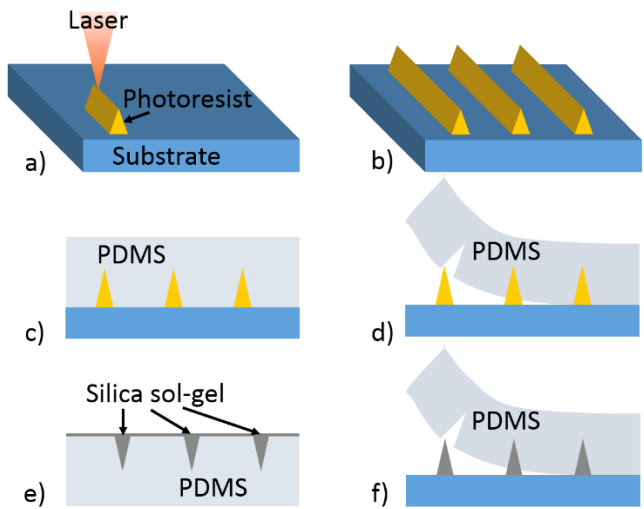

**Figure 1** Schematic description of the stamp making and printing process. a) First, the master sample is prepared by two-photon lithograpy, b) then it is developed and treated with chlorotrimethylsilane. c) PDMS is poured on the master sample and d) carefully pulled off after polymerization. e) The PDMS stamp is filled with sol-gel, placed on a substrate and f) removed after polymerization of the sol-gel.

Nanoscribe). An optional surface functionalization of (3-(N-ethylamino)isobutyl)trimethoxysilane (4wt% in methanol, Gelest SIE4886.0) for 30 minutes was seen to increase the reusability of the fabricated polymer master stamp by covalently bonding the resist to the silicon substrate. The structures were written in Nanoscribe's Galvo mode with a speed between 40000 and 70000 µm/s. After the writing process unpolymerized resist is removed by a 15-20 min bath in propylene glycol methyl ether acetate followed by 5 min in isopropanol. In order to condition the sample surface for stamp preparation the sample was left three hours in a box with a droplet of chlorotrimethylsilane. The polydimethylsiloxane (PDMS) was prepared by weighing 10 parts of SYLGARD® 184 silicone elastomer base and one part of the curing agent and subsequent mixing for 1 min and defoaming for 5 min. The PDMS was then poured on the master sample and captured air was removed with a vacuum treatment. Sample and PDMS were left in an oven at 80 ˚C for an hour in order to polymerize the PDMS. After cooling, the PDMS was carefully peeled from the master sample. The stamp was then filled with a silica sol-gel and placed on a substrate. Alternatively, the silica sol-gel can be drop casted or spin casted directly on the substrate and the stamp is placed on top afterwards. The stamp is left on the sample for at least three hours at room temperature to form a rigid silica sol-gel replica and is then peeled off. Residual sol-gel is left on the substrate for these demonstrations but is known to be on the order of ~15 nm, and can be removed via a quick dip in dilute HF with a minimal effect on resulting structures[13].

Residue-free printing of silver lines was achieved using capillary action. Stamps were cut open at their sides and put on top of the substrate such that good adhesion was achieved. A small amount of ink was brought in contact with the openings and pulled-in by capillary action. The distance that can be printed depends on the substrate surface and its treatment as well as on the stamp treatment. We obtain up to 5 mm long structures using silicon and indium tin oxide substrates by leaving the stamp in isopropanol for at least 4 hours. Furthermore, stamps can be filled from both sides in order to achieve twice this length.

## Results

Figure 2 shows scanning electron microscope (SEM) images of a) a master structure and b) a replica. The master sample was coated with 30 nm aluminum in order to facilitate SEM imaging the replica provided sufficient conductivity without coating. Six different geometries with different scales were explored: from left to right: micro lenses, micro volcanos, cones, pyramids, parabolic cross-section (micro concentrator) lines and triangular cross-section lines. The white scale bar corresponds to 100 µm.

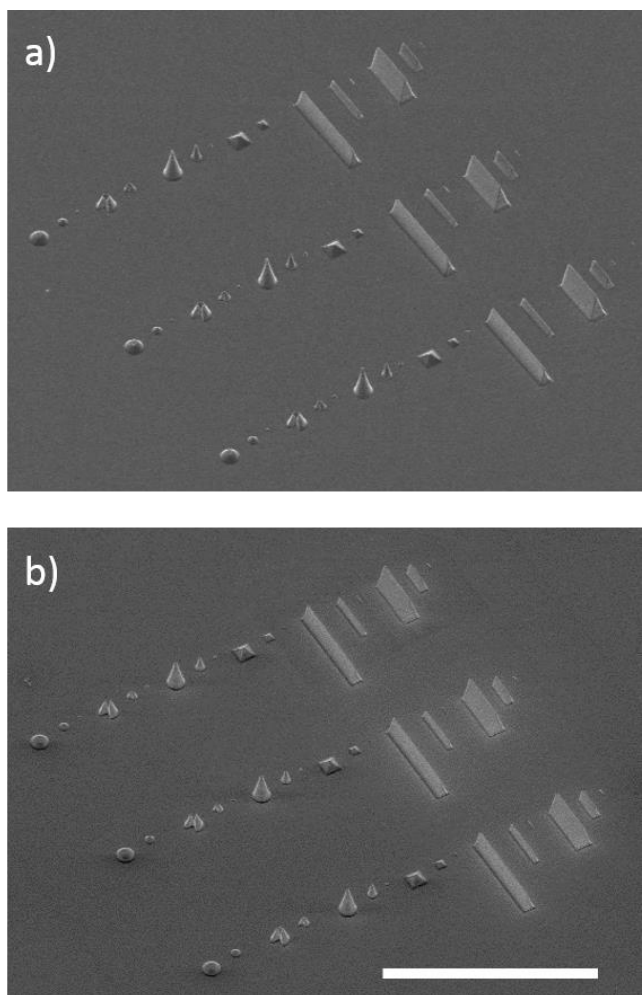

**Figure 2** SEM images taken under a 45˚ angle of a) the master structures that were prepared by two-photon lithography and b) printed replicas. The white scale bar corresponds to 100 µm.

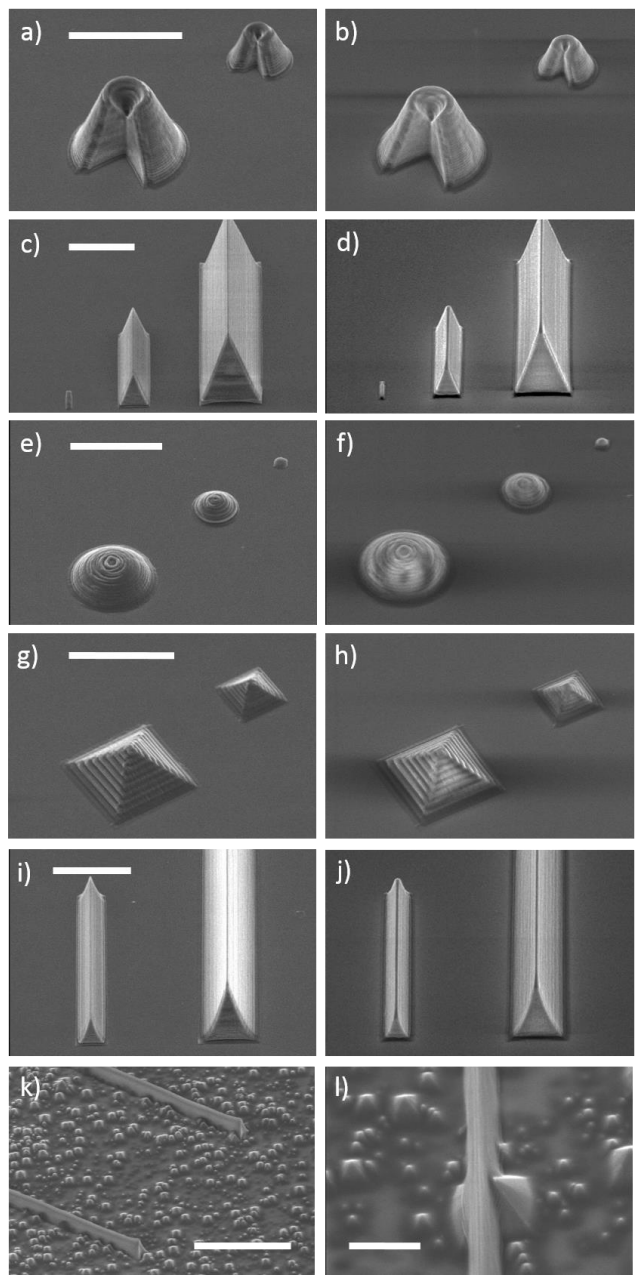

**Figure 3** SEM images taken under a 45˚ angle of a), b) and c) master structures and b), d) and f) replicas. The white scale bar corresponds to 10 µm. k) and l) show SEM images of triangular cross-section lines printed on structured silicon. Scale bar: k) 20 µm, l) 3 µm.

Figure 3 shows higher resolution images of the structures. a), c), e), g) and i) are master structures while b), d), f), h) and j) show the replicas. The white scale bars correspond to 10 µm. All shapes are reproduced with high fidelity. Moreover, even the sidewall textures are replicated showing that the quality of the printed shape is limited by the quality of the master structure.

Although the shapes chosen here all have a constant or monotonically increasing profile approaching the substrate it has been shown that structures with overhanging parts can be fabricated in this fashion as well[15].

Figure 3 k) and l) show SEM images of triangular cross-section lines printed on a textured silicon substrate. The printed structure experienced only small disturbance although this presents an extreme case with texture roughness being at similar dimensions as the printed features.

A known limitation of imprint lithography is the occurrence of residues in between printed features. While sol-gel residues are ~15 nm, and can be removed via a quick dip in dilute HF with a minimal effect on resulting structures[13], silver ink residues can lead to significant disturbance of the desired structure as shown in Figure 4c.

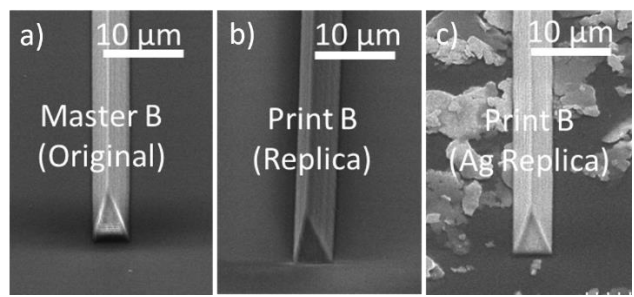

**Figure 4** Triangular cross-section lines a) master structure, b) printed with sol-gel and c) printed with silver ink.

As described in the Methods section it is possible to overcome this issue by first placing the stamp on the substrate and then infilling the lines by capillary action. SEM images of such silver lines are shown in Figure 5.

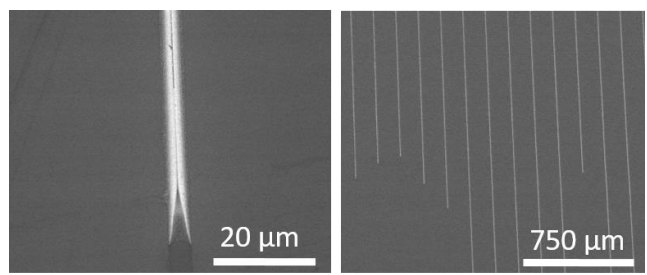

**Figure 5** Triangular cross-section lines printed with silver ink. Residue-free structures were achieved by infilling the PDMS stamp using capillary action.

## Conclusion

We have shown that 3-D nanostructures fabricated by two-photon lithography can be replicated by nanoimprint lithography. We have also demonstrated that 3-D structures can be transferred to non-standard substrates. Sub micrometer features are reproduced with high fidelity and the quality of the replicated structure exclusively depends on the master structure.

## AUTHOR INFORMATION

**Corresponding Author**

*rebsaive@caltech.edu

## ACKNOWLEDGMENTS

The information, data, or work presented herein was funded in part by the U.S. Department of Energy, Energy Efficiency and Renewable Energy Program, under Award Number DE-EE0006335 and by the Bay Area Photovoltaics Consortium under Award Number DE-

EE0004946. We thank Lucas Meza for two-photon lithography advice.